\documentclass[usenatbib]{mn2e}
\usepackage{amssymb}
\usepackage{graphicx}
\usepackage{subfigure}

\title[Photometry and spectroscopy of CS 1246]{Photometry and spectroscopy of the new sdBV CS 1246\footnote{Based on observations at the SOAR telescope, a collaboration between CPNq-Brazil, NOAO, UNC, and MSU.}}

\author[Barlow et al.]{B.N. Barlow$^{1}$$^{2}$\thanks{E-mail:bbarlow@physics.unc.edu}, B.H. Dunlap$^{1}$$^{2}$, J.C. Clemens$^{1}$, A.E. Lynas-Gray$^{3}$,
\newauthor K.M. Ivarsen$^{1}$, A.P. LaCluyze$^{1}$, D.E. Reichart$^{1}$, J.B. Haislip$^{1}$ \& M.C. Nysewander$^{4}$ \\
$^{1}$Department of Physics and Astronomy, University of North Carolina, Chapel Hill, NC 27599-3255, USA\\
$^{2}$Visiting astronomer, CTIO, NOAO, which is operated by the AURA, under contract with the NSF.\\
$^{3}$Department of Physics, University of Oxford, Keble Road, Oxford OX1 3RH, England\\
$^{4}$Alion Science \& Technology, 1000 Park Forty Plaza, Durham, NC 27713, USA}\begin{document}

\maketitle

\label{firstpage}
\begin{abstract}
We report the discovery of a large-amplitude oscillation in the hot subdwarf B star CS 1246 and present multi-colour photometry and time-resolved spectroscopy supporting this discovery.  We used the 0.41-m PROMPT array to acquire data in the u', g', r', and i' filters simultaneously over 3 consecutive nights in 2009 April.  These data reveal a single oscillation mode with a period of 371.707 ${\pm}$ 0.002 s and an amplitude dependent upon wavelength, with a value of 34.5 ${\pm}$ 1.6 mma in the u' filter.  We detected no additional frequencies in any of the light curves.  Subsequently, we used the 4.1-m SOAR telescope to acquire a time-series of 248 low-resolution spectra spanning 6 hrs to look for line profile variations.  Models fits to the spectra give mean atmospheric values of T$_{eff}$ = 28450 $\pm$ 700 K and log \textit{g} = 5.46 $\pm$ 0.11 undergoing variations with semi-amplitudes of 507 $\pm$ 55 K and 0.034 $\pm$ 0.009, respectively.  We also detect a radial velocity oscillation with an amplitude of 8.8 $\pm$ 1.1 km s$^{-1}$.  The relationship between the angular and physical radii variations shows the oscillation is consistent with a radial mode.  Under the assumption of a radial pulsation, we compute the stellar distance, radius, and mass as d = 460 $\pm$ $^{190}_{90}$ pc, R = 0.19 $\pm$ 0.08 R$_{\sun}$, and M = 0.39 $\pm^{0.30}_{0.13}$ M$_{\sun}$, respectively, using the Baade-Wesselink method.
\end{abstract}


\section{Introduction}

Hot subdwarf B (sdB) stars are evolved objects presumed to have He-burning cores and modelled as extended horizontal branch stars \citep{heb86}.  They dominate surveys of faint blue objects and are often cited as the main source of the UV-upturn observed in globular clusters and giant elliptical galaxies.  Their temperatures range from 20000 to 40000 K, and their log \textit{g} values span more than an order of magnitude from 5 to 6.2.  Their evolutionary histories are not entirely understood; various formation scenarios are possible including both single-star and binary-star channels \citep{dcr96,han02,han03}.  At least two thirds of them appear to be in binary systems \citep{saf01,max01}, but a higher binary fraction cannot be ruled out.  \citet{han02,han03} have even shown it is possible to explain 100\% of the sdB stars as the products of binary-star channels.  In either case, sdB stars probably evolve from the red giant branch after losing a signficant portion of their outer H envelopes, and stellar models show they will evolve directly to the white dwarf cooling sequence after exhausting the He in their cores \citep{dor93}.  Their optical spectra are dominated by H Balmer lines and sometimes also display He I or even the 4686 \AA\ He II line.  See Heber (2009) for a review on what is currently known about sdB stars.

More than a decade ago, \citet{kil97} reported the first detection of photometric oscillations in a sdB star (EC 14026-2647).  Contemporaneous with this discovery, \citet{cha96,cha97} predicted the existence of a class of sdB pulsators after calculating stellar models that showed driving of p-modes.  The oscillations they found were driven by the $\kappa$-mechanism and associated with an opacity bump from the ionisation of Fe in their model envelopes.  The discovery of the first pulsator sparked an interest in sdB variability studies, and, subsequently, around 70 pulsators have been discovered to date.  Many techniques have been applied to sdB pulsators successfully including multi-colour photometry (see \citealt{tre06} for a summary) and time-resolved spectroscopy (pioneered by \citealt{oto00,tel04,til07}).  Asteroseismological studies of these stars have revealed information about their internal structure, including total mass, envelope mass, and chemical stratification and have been carried out on a handful of sdB pulsators (see \citealt{bra01} and \citealt{cha05} as examples).  \citet{ost09} presents a thorough review of asteroseismolgy and its successes.

Observationally, the subdwarf B pulsators may be divided into three classes: the rapid (sdBV$_{r}$), slow (sdBV$_{s}$), and hybrid (sdBV$_{rs}$) pulsators\footnote{Here we adopt the naming scheme in which all sdB pulsators are given the base name 'sdBV'; subscripts 'r' and 's' are added upon observations of rapid and slow oscillations, respectively.  For reference, the sdBV$_{r}$ stars are also called V361 Hya and EC 14026 stars, while the sdBV$_{s}$ stars have the aliases PG 1716, V1093 Her, and 'Betsy' stars.}.  The sdBV$_{r}$ stars are associated with models of p-mode pulsators and have periods between 80 and 600 s and typical amplitudes less than 1\%.  In the log \textit{g}-T$_{eff}$ plane, they cluster together in an instability strip with T$_{eff}$ ranging from 28000 to 35000 K and log \textit{g} between 5.2 and 6.1.  The sdBV$_{s}$ stars generally have periods from 1 to 2 hrs with amplitudes of only a few tenths of a percent; their oscillations are modelled with g modes.  They are also found in an instability strip with T$_{eff}$ between 23000 and 30000 K and log \textit{g} near 5.4.  In the region where the blue edge of the sdBV$_{s}$ instability strip overlaps the red edge of the sdBV$_{r}$ strip, three sdBV$_{rs}$ stars have been observed with fast and slow oscillations \citep{sch06,bar06,lut09}.  Both of the instability strips contain stars that have been observed not to vary photometrically.

The most fundamental parameters of any star, the mass and radius, are not well known for sdB stars and are rarely computed from observations.  Single and binary evolution calculations point to a canonical mass of 0.47 M$_{\sun}$ \citep{han02,han03}, and a few masses supporting this result have been derived from asteroseismology (e.g., \citealt{bra01,ran07}) and from eclipsing binaries (e.g., \citealt{woo93}, \citealt{dre01}).  \citet{zha09} have even proposed a new technique for deriving statistical masses from observational data in the literature.  Most of the sdB radii measurements to date come from binary systems for which the separation distance is known; the sdB radius is derived from the ratio of the radii and assumes information about the companion.  Such studies show a typical sdB radius to be near 0.2 R$_{\sun}$, but additional and more direct measurements are needed to confirm this result. 

We report the detection of a single, large-amplitude oscillation in the sdB star CS 1246 ($\alpha _{2000}$=12:49:37.69, $\delta _{2000}$= -63:32:08.99), an object first observed by \citet{rei88} in a survey of hot white dwarfs and OB subdwarfs in obscured regions of the Galactic plane.  In one of the most opaque sections of the Coalsack Dark Nebula, CS 1246 was identified by \citet{rei88} as a candidate hot subdwarf B star with a V magnitude of 14.59 $\pm$ 0.14 and U-B and B-V colours of -1.08 $\pm$ 0.10 and 0.28 $\pm$ 0.10, respectively.  Neither T$_{eff}$ nor log \textit{g} has previously been determined.

We first learned of CS 1246 while browsing the online Subdwarf Database \citep{ost06} for candidate pulsators observable with the 0.41-m Panchromatic Robotic Optical Monitoring and Polarimetry Telescopes (PROMPT) in Chile.  Even though there were no available estimates of T$_{eff}$ or log \textit{g}, we decided to look for pulsations since the U-B colour roughly matched that of known pulsators and the star was a brighter, Southern-hemisphere target.  The discovery run data revealed an interesting light curve, and we subsequently obtained multi-colour time-series photometry and time-resolved spectroscopy to study the pulsations.  We observed line profile variations in the spectra and found substantial oscillations in the radial velocity, temperature, and gravity.  The nature of these variations is consistent with that expected for an $\ell = 0$ pulsation, but we cannot rule out the possibility of non-radial modes with our current data set.  Although not one of our original goals, we were able to compute the distance, radius, and mass of CS 1246 from the relation between the physical and angular radii variations.  Our study is the first to apply such a method to an sdBV star, and the results show the mass and radius to be consistent with those calculated for other sdB stars in binaries and from theory.

\section{DISCOVERY OF THE PULSATION}

\begin{figure}
  \centering
  \includegraphics{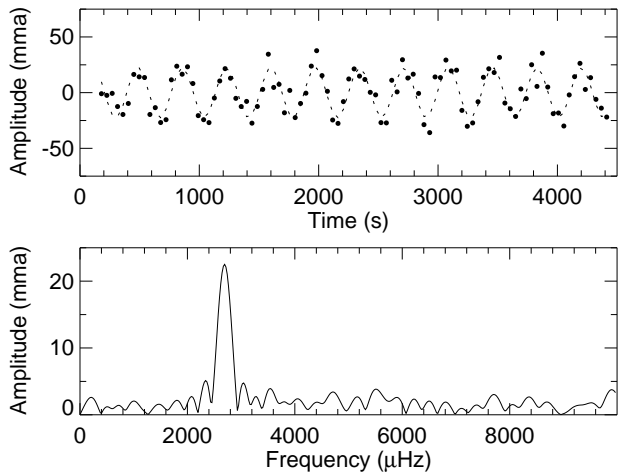}
   \caption{Light curve (upper panel) and amplitude spectrum (lower panel) of the discovery run data from 2009 March 30.  The light curve was produced from V-filtered data and is presented unsmoothed.} 
  \label{fig:discovery}
\end{figure}

We first observed CS 1246 on 2009 March 30 after submitting a job to the PROMPT queue using the web-based SKYNET interface (see \S \ref{phot:instrument} for a description of these systems).  Later that night PROMPT 4 monitored the star for approximately 75 min, obtaining an uninterrupted series of 80 exposures through a Johnson V filter with integration times of 40 s and a duty cycle of 89\%.  Even before removing atmospheric extinction effects or transparency variations, we realized CS 1246 was a pulsator as its oscillations were visible in the undivided, raw light curve.  After formally reducing and analysing the data using the methods described in \S \ref{phot:reductions} and \S \ref{phot:analysis}, we found a single oscillation with a period of 372 s and an amplitude of 23 mma.  Figure \ref{fig:discovery} shows the reduced light curve and its amplitude spectrum.  No other frequencies were observed in this data set, which had a mean noise level of 1.6 mma.  The relatively long period and large amplitude of the oscillation inspired us to obtain simultaneous multi-colour photometry with the PROMPT and time-resolved spectroscopy with the SOAR telescope.  These observations and the results thereof are described in the sections that follow.

\section{MULTI-COLOUR PHOTOMETRY}
\subsection{The instrument}
\label{phot:instrument}
PROMPT is a system of 0.41-m Ritchey-Chr\'etien telescopes at Cerro Tololo Inter-American Observatory (CTIO) in Chile built primarily for rapid and simultaneous multiwavelength observations of gamma-ray burst afterglows.  Table 1 presents some of the CCD characteristics of these telescopes.  The PROMPT telescopes are 100\% automated and operate under the control of SKYNET, a prioritised queue scheduling system that allows registered users to acquire, view, and download data from any location via a PHP-enabled Web server.  Observations in the queue are assigned priority levels and scheduled accordingly; gamma-ray bursts receive top priority.  The SKYNET system enables automated data collection at times that would be inconvenient or impossible for a human observer.  For example, our multi-colour photometric programme was executed by PROMPT while BNB and BHD were in Detroit, Michigan, supporting the University of North Carolina Men's Basketball team at the 2009 NCAA National Championship.  Obtaining the data presented in \S \ref{phot:reductions} using classical observing methods would have been an expensive and time-consuming undertaking.  For additional details on PROMPT and SKYNET, refer to Reichart et al. (2005).

\begin{table}
\caption{Characteristics of the PROMPT CCDs}
\begin{tabular}{cccccc}
\hline
Telescope & Size & Pixel Scale &Gain & Readnoise\\
 & (pix) & (arcsec/pix) & (e${^-}$/ADU) & (e${^-}$)\\
\hline
PROMPT 2 & 3k x 2k & 0.41 & 1.6 & 8.3\\
PROMPT 3 & 1k x 1k & 0.59 & 1.6 & 8.3\\
PROMPT 4 & 1k x 1k & 0.59 & 1.4 & 7.6\\
PROMPT 5 & 1k x 1k & 0.59 & 1.4 & 9.2\\
\hline
\end{tabular}
\label{table:obs}
\end{table}

\begin{figure}
  \centering
  \includegraphics[width=200pt]{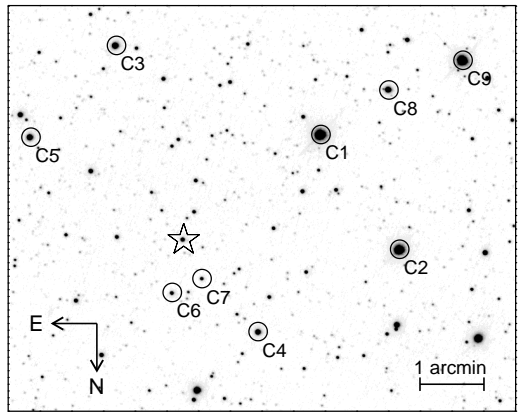}
  \caption{CS 1246 field.  Image shown is a stack of 30 40-s V-band frames taken with PROMPT 4.  CS 1246 is indicated with a star while the comparison stars are marked with circles and labeled C1-C9.} 
  \label{fig:field}
\end{figure}

\subsection{Observations and reductions}
\label{phot:reductions}
PROMPT obtained multi-colour photometry simultaneously through the u', g', r', and i' filters on 2009 April 4, 5, and 6 using all available telescopes.  Table \ref{table:phot:log}  presents a detailed log of these observations.  The CCD parameters of the PROMPT telescopes (Table 1) and SKYNET overhead resulted in duty cycles of 94\%, 80\%, 89\%, and 89\% for the u', g', r', and i' filter data sets, respectively, during the continuous portions of the runs.  Although the majority of the runs were uninterrupted, each data set contains an approximately 15-minute long gap due to the inability of the equitorial mounts to track through the meridian.  Additional 40-minute gaps are present in the u' and g' data from 2009 April 6.  In total, our 3-day photometry campaign produced 113 hours of data with 7291 useable frames.

\begin{table*}
\centering
\caption{Multicolour photometry observations log}
\begin{tabular}{ccccccccc}
\hline
Start Time$^{a}$ & Date & Filter & Telescope & T$_{exp}$ & T$_{cycle}$ & Length & \# Frames & Comparison Stars\\
(UT) & (UT) & & & (s) & (s) & (hrs) & & \\
\hline
23:59:02.8 & 2009 Apr 04 & u' & PROMPT 3 & 80 & 86 & 9.4 & 393 & C1,C2,C9\\
00:57:03.3 & 2009 Apr 05 & g' & PROMPT 2 & 40 & 52 & 8.7 & 592 & C1,C2,C3,C4,C8,C9\\
00:01:47.7 & 2009 Apr 05 & r' & PROMPT 4 & 40 & 47 & 9.6 & 710 & C2,C3,C4,C5,C7\\
00:01:51.7 & 2009 Apr 05 & i' & PROMPT 5 & 40 & 47 & 9.6 & 710 & C2,C3,C4,C5,C6\\
\hline
23:57:56.0 & 2009 Apr 05 & u' & PROMPT 3 & 80 & 86 & 9.4 & 380 & C1,C2,C9\\
23:57:05.9 & 2009 Apr 05 & g' & PROMPT 2 & 40 & 52 & 8.6 & 580 & C1,C2,C3,C4,C8,C9\\
23:57:41.8 & 2009 Apr 05 & r' & PROMPT 4 & 40 & 47 & 9.3 & 700 & C2,C3,C4,C5,C7\\
23:57:38.9 & 2009 Apr 05 & i' & PROMPT 5 & 40 & 47 & 9.3 & 700 & C2,C3,C4,C5,C6\\
\hline
23:56:41.8 & 2009 Apr 06 & u' & PROMPT 3 & 80 & 86 & 9.7 & 395 & C1,C2,C9\\
23:55:47.2 & 2009 Apr 06 & g' & PROMPT 2 & 40 & 52 & 9.7 & 663 & C1,C2,C3,C4,C8,C9 \\
23:56:36.1 & 2009 Apr 06 & r' & PROMPT 4 & 40 & 47 & 9.7 & 735 & C2,C3,C4,C5,C7 \\
23:56:31.8 & 2009 Apr 06 & i' & PROMPT 5 & 40 & 47 & 9.7 & 733 & C2,C3,C4,C5,C6\\
\hline
\multicolumn{9}{l}{\footnotesize{$^{a}$Time at midpoint of first exposure.}}\\

\end{tabular}
\label{table:phot:log}
\end{table*}

\begin{figure*}
  \centering
 \includegraphics{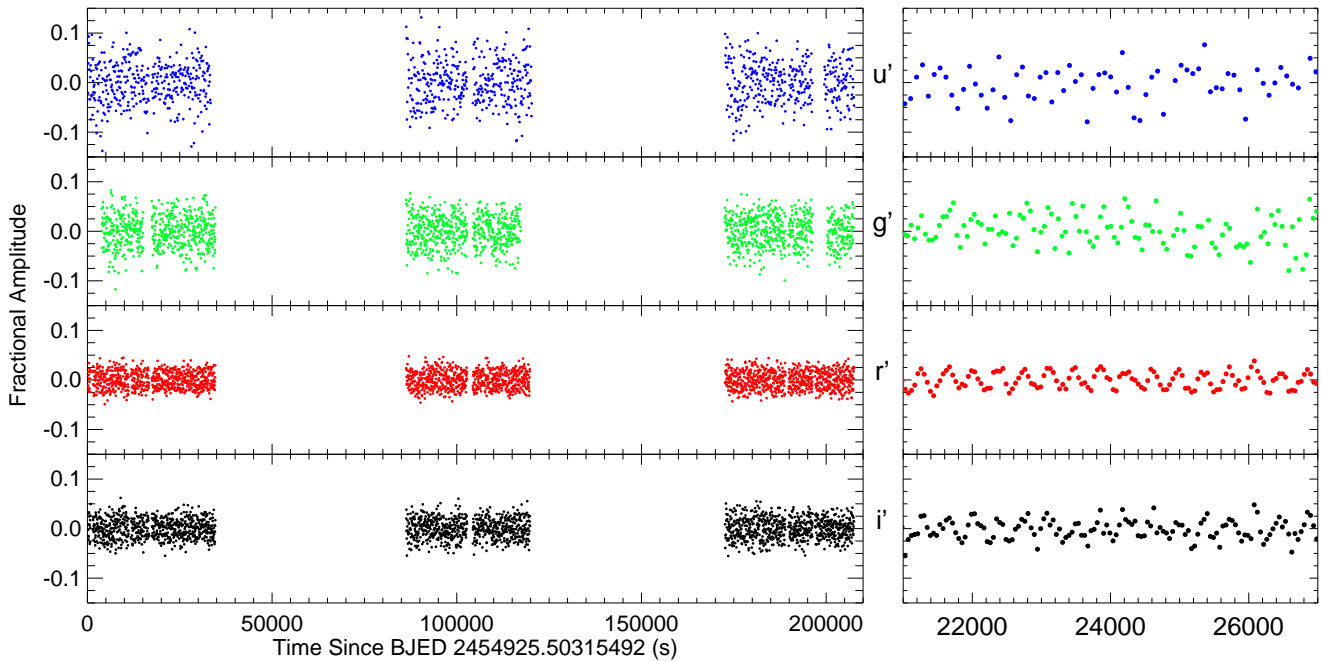}
   \caption{Light curves of CS 1246 over 3 consecutive nights (left panels) taken simultaneously through the u', g', r', and i' filters (from top to bottom).  These observations represent a total of 113 hours of data taken with 7291 frames.  Expanded views (right panels) of the same 1.5-hour section of the curves are shown to illustrate the variable nature of the star.} 
  \label{fig:phot:lcs}
\end{figure*}

After bias-subtracting, dark-subtracting, and flat-fielding the frames with IRAF using conventional procedures, we extracted our photometry using the external IRAF package CCD\_HSP written by Antonio Kanaan.  Aperture radii were chosen to maximize the signal-to-noise (S/N) ratio in the light curves and were always less than the seeing width.  We used sky annuli to subtract the sky counts from each stellar aperture, being sure to avoid contamination of the annuli with other stars in the field.  We divided the sky-subtracted light curves by an average of constant comparison star light curves to correct for transparency variations; these stars are labeled in the field image shown in Figure \ref{fig:field}.  Some of them lacked an adequate S/N ratio through some filters while saturating the CCD in others, and, consequently, we employed different comparison stars for different filters (see Table \ref{table:phot:log}).  We are confident that the results of our frequency analyses are independent of the choices of the comparison stars, as they all appeared to be constant.  Atmospheric extinction effects were approximated and removed each night by fitting and normalising the curves with parabolas.  We used the WQED suite \citep{tho09} to analyze the data and apply timing corrections to the barycentre of the solar system.  Figure \ref{fig:phot:lcs} presents the full, differentially-corrected light curves.

\subsection{Frequency analysis}
\label{phot:analysis}
\begin{figure*}
  \centering
  \subfigure(a)
  {
     \includegraphics{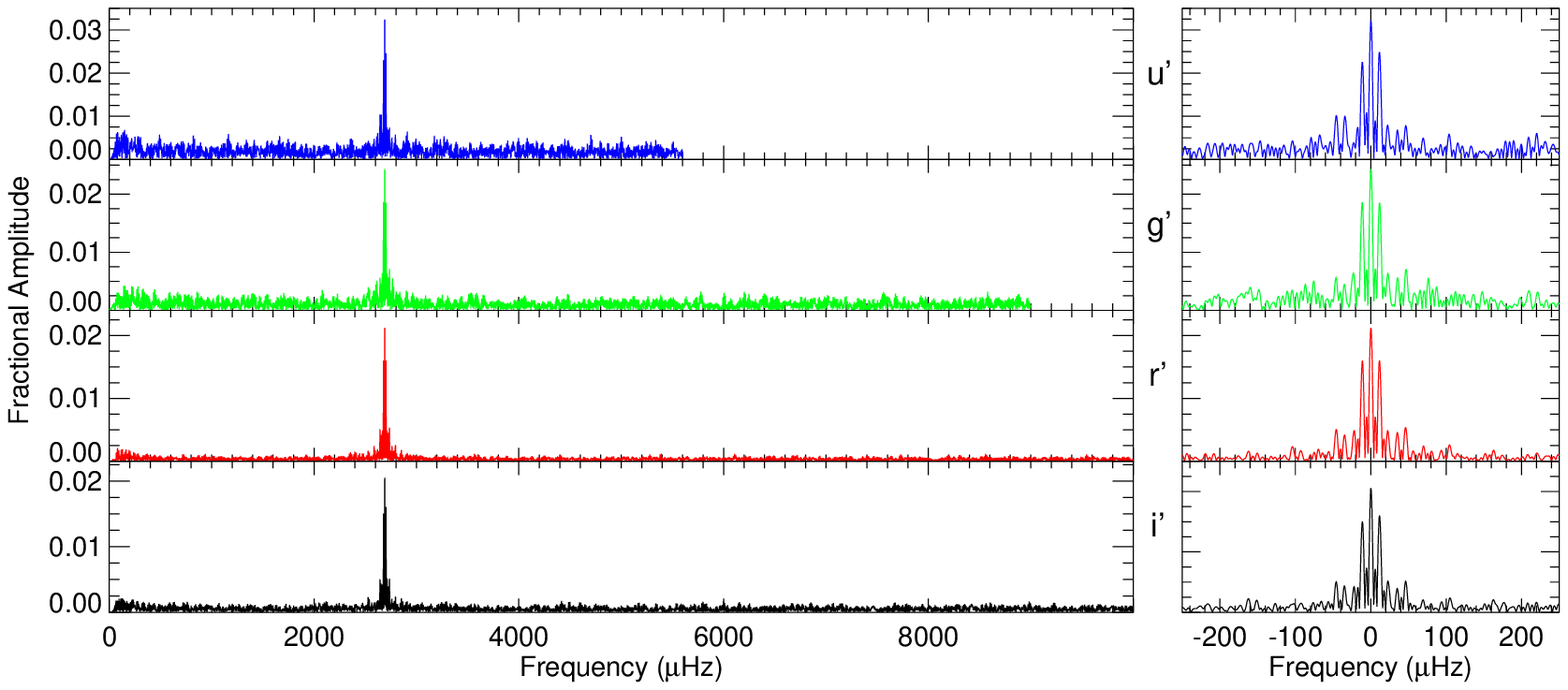}
     
  }
 
  \subfigure(b)
  {
     \includegraphics{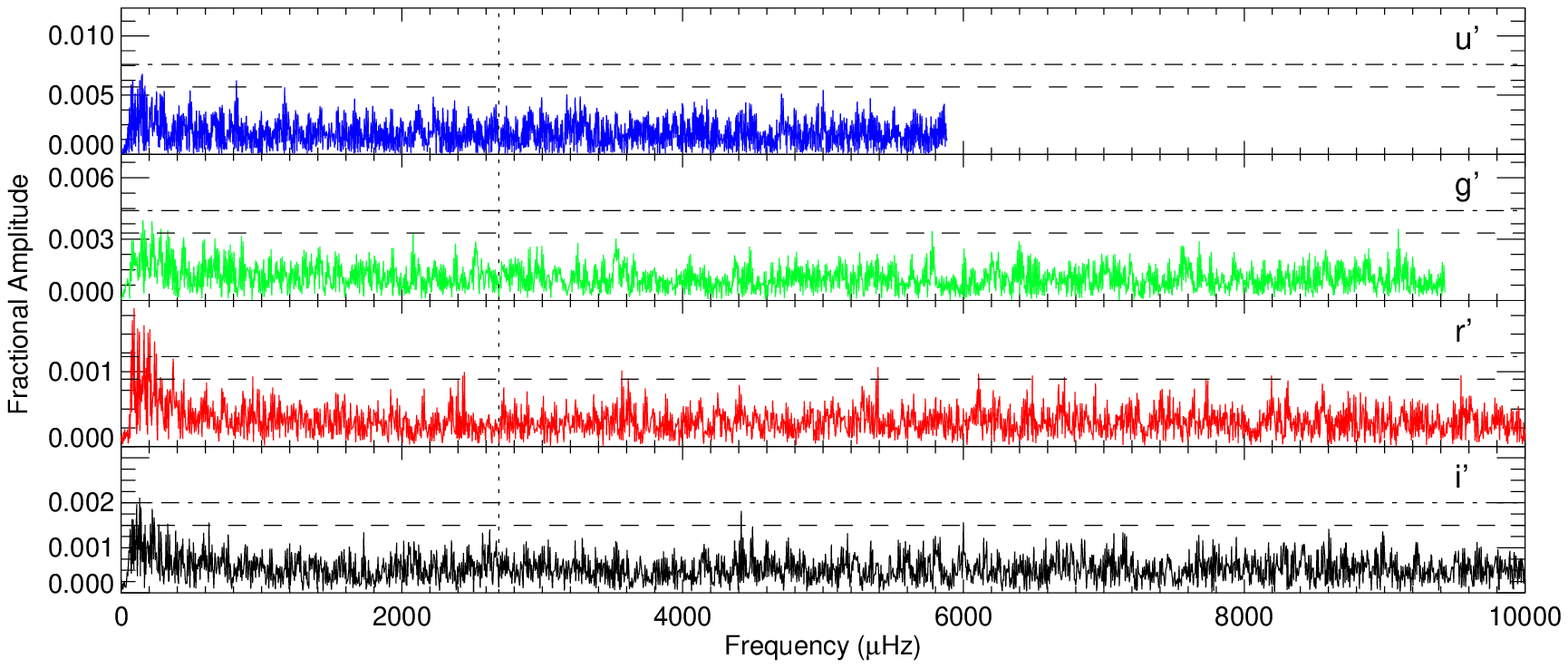}
     
  }
  
  \caption{Amplitude spectra of the combined light curves in the u', g', r', and i' filters (from top to bottom) before (a) and after (b) pre-whitening of the main mode.  The window functions for each data set are shown in the right portion of (a).  In the pre-whitened spectra, the 3$\sigma$ and 4$\sigma$ levels are represented by dashed and dot-dashed lines, respectively, and the position of the pre-whitened signal is marked with a vertical dashed line.  The u' and g' spectra are plotted out to their respective Nyquist frequencies.  The r' and i' plots are terminated at 10 mHz but are consistent with noise beyond this point.} 
  \label{fig:phot:fts}
\end{figure*}

We analysed our reduced light curves using two tools:  the discrete Fourier transform and the least-squares fitting of sine waves.  WQED, again, was used to access these tools.  Figure \ref{fig:phot:fts}a presents the amplitude spectra for the combined light curves shown in Figure \ref{fig:phot:lcs} along with their window functions.  The u' and g' amplitude spectra are plotted out to their respective Nyquist frequencies, while the r' and i' spectra are terminated at 10 mHz (beyond this point these spectra are consistent with noise). 

Inspection of the amplitude spectra reveals a single, dominant mode at 2690.29 $\mu$Hz (371.707s) with amplitudes of 32.0 $\pm$ 1.5, 24.4 $\pm$ 0.9, 21.3 $\pm$ 0.3, and 20.6 $\pm$ 0.4 mma in the u', g', r', and i' passbands, respectively.  Table \ref{table:phot:fits} summarizes the results of the least-squares fits; the errors shown are derived directly from the fits and assume uncorrelated noise.  The results of \citet{mon99} imply these errors may be larger by a factor of three. We detect no phase differences between the four colour curves within our level of precision.  As the integration times of our exposures are significant fractions of the 371.7-s period, we must take phase smearing into account.  For any measured observable oscillating with a period $P$ and a physical amplitude $A$, a finite integration time $I$ will result in a reduced observed amplitude $A_{o}$ given by \begin{equation} A_{o}=\frac{P}{\pi I} sin(\frac{\pi I}{P})A, \end{equation}  \citet{bal99} presents a discussion of phase smearing and a derivation of this equation.  The amplitudes we measure for the 371.7-s variation underestimate the physical amplitudes by 7.4\% in the u' filter and 1.9\% in the g', r', and i' filters.  The smearing-corrected amplitudes are 34.5 $\pm$ 1.6, 24.8 $\pm$ 0.9, 21.7 $\pm$ 0.3, and 21.0 $\pm$ 0.4 mma in the u', g', r', and i' filters, respectively.   

After fitting the dominant mode and subtracting the fit from the light curves, we recalculated the Fourier transforms of the residual light curves; they are shown in Figure \ref{fig:phot:fts}b.  The mean noise levels in these pre-whitened u', g', r', and i' amplitude spectra are 1.9, 1.1, 0.3, and 0.5 mma, respectively.  All of the spectra show a heightened noise level at extremely low frequencies, but this is likely a side effect of our method for extinction removal.  Beyond the low-frequency regime, some of the pre-whitened spectra show peaks at or above the 3-$\sigma$ level.  However, none of these periodicities were detected in more than one data set, whether from night-to-night or from filter-to-filter, so we evaluate their signficance using the false alarm probability \citep{hor86}.  None of them meet our significance test of a false alarm probability $<$1\%.  Moreover, the folded pulse shapes agree with that of a single sinusoid to within the errors, implying a lack of harmonically-related frequencies in the light curve.

\begin{figure}
  \centering
  \includegraphics{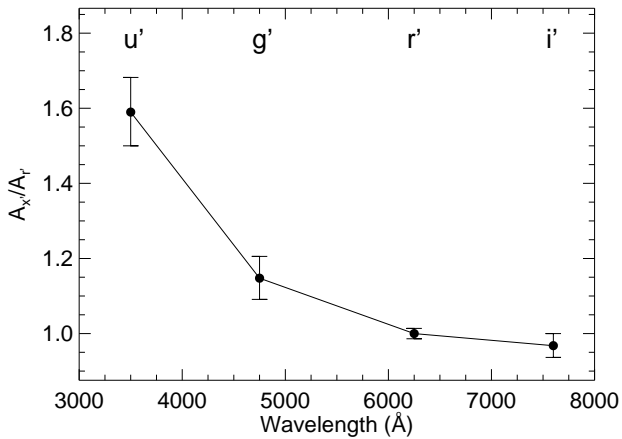}
  \caption{Amplitude ratios A$_{x'}$/A$_{r'}$ of the 371.707s mode for the u', g', r', and i' data, corrected for phase-smearing.  Normalisation to A$_{r'}$ was chosen to minimize the error bars.} 
  \label{fig:spec:amplitude_ratios}
\end{figure}

The amplitude of the oscillation is strongly wavelength-dependent, as illustrated in Figure \ref{fig:spec:amplitude_ratios}.  The u' amplitude is the largest, being 39\%, 59\%, and 64\% greater than those in the g', r', and i' filters, respectively.  In principle, these ratios may be used along with the known atmospheric parameters of the star to estimate the degree \textit{l} of the oscillation, but in our case the large error bars make such an exercise inconclusive.  

\begin{table}
\centering
\caption{Results of the least-squares fits to the light curves}
\begin{tabular}{cccc}
\hline
Filter & Frequency & Amplitude$^{b}$ & Phase$^{c}$\\
 & ($\mu$Hz) & (mma) & (s) \\
\hline
u' & 2690.28 $\pm$ 0.10 &  32.0 $\pm$ 1.5 & 76.5 $\pm$ 2.8\\
g' & 2690.30 $\pm$ 0.08 &  24.4 $\pm$ 0.9 & 75.7 $\pm$ 2.2\\
r' & 2690.29 $\pm$ 0.03 &  21.3 $\pm$ 0.3 & 77.3 $\pm$ 0.8\\
i' & 2690.30 $\pm$ 0.05 &  20.6 $\pm$ 0.4 & 77.5 $\pm$ 1.2\\
\hline
\multicolumn{4}{l}{\footnotesize{$^{b}$ Not corrected for phase-smearing.}}\\
\multicolumn{4}{l}{\footnotesize{$^{c}$ Time of maximum measured from BJED 2454925.50315492.}}\\
\end{tabular}
\label{table:phot:fits}
\end{table}

\section{TIME-RESOLVED SPECTROSCOPY}
\label{spec}
\subsection{The instrument}
The Goodman Spectrograph is an imaging spectrograph built for one of the Nasmyth ports of the 4.1-m SOAR telescope on Cerro Pachon in Chile.  In imaging mode, the camera-collimator combination re-images the SOAR focal plane with a focal reduction of three times, yielding a plate scale of 0.15 arcsec pixel$^{-1}$ at the CCD.  The camera contains a 4k x 4k Fairchild 486 back-illuminated CCD with electronics and dewar provided by Spectral Instruments, Inc. of Tucson, Arizona.  The system uses optics of fused silica, NaCl, and CaF$_{2}$ to achieve high throughput from 320 to 850 nm.  In spectroscopic mode, the instrument can obtain spectra of objects over a 3\arcmin x 5\arcmin field using long slits with widths ranging from 0.46\arcsec\ to 10\arcsec.  A multi-slit mode is currently being commissioned.  Users may choose from three Volume Phase Holographic (VPH) gratings with dispersions of 0.31, 0.65, and 1.3 \AA\ pixel $^{-1}$.  For further details concerning the spectrograph, refer to \citet{cle04}.

\subsection{Observations and Reductions}
We obtained 248 low-resolution spectra of CS 1246 during an engineering run on 2009 April 17 with the Goodman Spectrograph in order to look for line profile variations.  The 600 mm$^{-1}$ VPH grating (0.65 \AA\ pixel$^{-1}$ dispersion) was used to cover a spectral range from 3500 to 6200 \AA.  As the sky at Cerro Pachon was photometric with reasonable seeing, we employed a 10\arcsec-wide slit and let the average seeing of 1.1\arcsec\ determine our final resolution of 4.8 \AA.  The position angle was set to 272.3 degrees E of N so that we could place a nearby comparison star on the slit.  We binned the spectral images by two in the spatial and dispersion directions and only read out 400 binned pixels along the slit to minimize the readout time.  Each exposure had an integration time of 80 s with 6 s of overhead between images, resulting in a duty cycle of 93\%.  The entire run lasted 6 hours from 00:59:42.0 to 07:07:53.4 UT with a 15-minute gap that began near 02:19:00 UT.  Comparison spectra of HgAr and Cu lamps were obtained for wavelength calibration during the gap and after the entire series was complete.

\begin{figure}
  \centering
  \includegraphics[0in,0in][3.3in,3.1in]{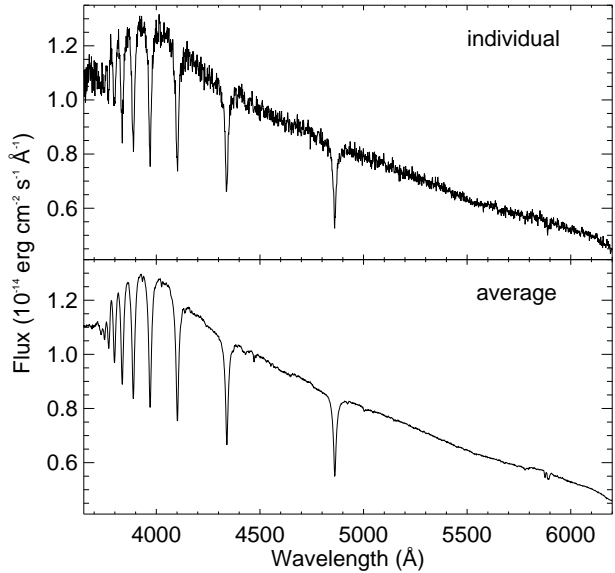}
  \caption{Individual (top) and average (bottom) spectra of CS 1246 taken with the Goodman spectrograph on 2009 Apr 17.  The spectra shown were flux-calibrated using the spectrophotometric standard EG 274.} 
  \label{fig:spectra}
\end{figure}

We bias-subtracted and flat-fielded our spectra using standard tasks within IRAF.  The \textit{apall} routine was used to optimally extract one-dimensional spectra from the frames and subtract a fit to the sky background.  The resulting individual spectra have an average S/N ratio of 49 pixel$^{-1}$ in the continuum near 5000 \AA.  The top panel of Figure \ref{fig:spectra} shows an example of one of these spectra.  Unfortunately, we were not succesfully able to use the HgAr or Cu frames to wavelength-calibrate our spectra, as most of the 43-\AA\ wide lines overlapped and made it difficult to accurately mark line centres.  As an alternative, we self-calibrated the mean spectrum to the H Balmer and He I lines and used the solution to wavelength-calibrate the individual spectra.  Although this technique makes it impossible to measure the true space velocity of CS 1246, it preserves the relative wavelength scale without removing offsets between spectra and does not effect our relative radial velocity measurements.  We checked the resulting wavelength solution with that derived from past spectra taken with the same instrumental setup (but with smaller slits) and found that the dispersions agreed well.  

\subsection{The mean spectrum}
\label{mean_spectrum}
The average spectrum has a S/N ratio of 770 pixel$^{-1}$ and is shown in the bottom half of Figure \ref{fig:spectra}.  H Balmer (H$\beta$-H13) and He I lines are the most prominent features in the spectrum, but signatures of Si III, N II, and O II are also present.  The spectrum appears to be contaminated with Ca II (3933 \AA) and Na D.  Although these features usually indicate the presence of a cooler companion, we expect some level of interstellar absorption given the extremely low galactic latitude (-0.94 $\deg$) of CS 1246.  Moreover, both Ca and Na D have been observed in the spectra of other stars located in and behind the Coalsack \citep{fra00,fra89a,cra91}.  No other spectral features indicative of a companion are present, and we currently have no reason to believe the lines originate from a companion star.  

To determine the atmospheric parameters from the mean spectrum, we employed the model grid of synthetic optical spectra described in \citet{han07}.  This grid is defined in terms of 21 values of T$_{eff}$ from 20000K to 40000K in increments of 1000K, 11 values for the surface gravity from log \textit{g} of 5.0 to 7.0 in increments of 0.2, and 4 values of the helium-to-hydrogen ratio from log \textit{N}(He)/\textit{N}(H) of -3 to 0 in increments of 1.  To obtain further resolution in the grid, we interpolated the models using a cubic spline interpolation technique (see \citealt{pre07}).  Each model spectrum was degraded by convolving it with a Gaussian whose FWHM mimicked that of a single resolution element in our spectrum.  Finally, we re-binned the model spectra to match the binning of the observed spectrum.

The comparison to the model spectra needed to be carried out in terms of normalised line profiles, and so we implemented a fitting technique similar to that described in \citet{ber95}.  We normalised the flux in the He I and H Balmer (H$\beta$-H9) profiles to a linear continuum set to unity at two points.  These points are located at fixed distances on either side of the profile and are far enough from the line centre to completely encompass the absorption profile.  The unity point fluxes can be determined by averaging the fluxes in the neighbouring pixels, but defects in the observed spectrum at these locations may lead to inaccurate fits to the continuum.  Instead, we fit each absorption line and its surrounding continuum with a ``pseudo'' Lorentzian profile as described in \citet{saf88}.  This function has the form \begin{equation} G(\lambda) = C_{0} + C_{1}\lambda + C_{2}\lambda^{2} + \frac{A}{1+[(\lambda-\mu)/\sqrt{2}\sigma]^{\alpha}}. \end{equation} The exponent $\alpha$ controls the overall shape of the function and fits both the peak and the wing of an absorption line much better than a classic Lorenztian or Gaussian profile.  We used the IDL routine MPFIT \citep{mar09}, which employs the Levenberg-Marquardt method \citep{pre07}, to perform a non-linear least-squares fit to each line.  The unity point fluxes are derived from the resulting smooth fit, and each profile is normalised by dividing by the straight line passing through these points.  We normalised the line profiles in model spectra in the same manner except that the unity point fluxes were taken directly from the spectra (which are noiseless) instead of being determined by a smooth fit.  

We determined the best-fit model for each line profile with a $\chi^{2}$ minimization technique and found T$_{eff}$ = 28450 $\pm$ 300, log \textit{g} = 5.46 $\pm$ 0.06, and log \textit{N}(He)/\textit{N}(H) = -2.0 $\pm$ 0.1.  Systematic errors in the data and from fitting to imperfect models also contribute to the uncertainties in these values.  Consequently, we believe the statistical errors presented above underestimate the uncertainties and adopt the following values:  T$_{eff}$ = 28450 $\pm$ 700, log \textit{g} = 5.46 $\pm$ 0.11, and log \textit{N}(He)/\textit{N}(H) = -2.0 $\pm$ 0.3.  These parameters place CS 1246 near the boundary of the sdBV$_{r}$ and sdBV$_{s}$ instability strips.

\subsection{Spectrophotometric light curve}
\begin{figure}
  \centering
  \includegraphics{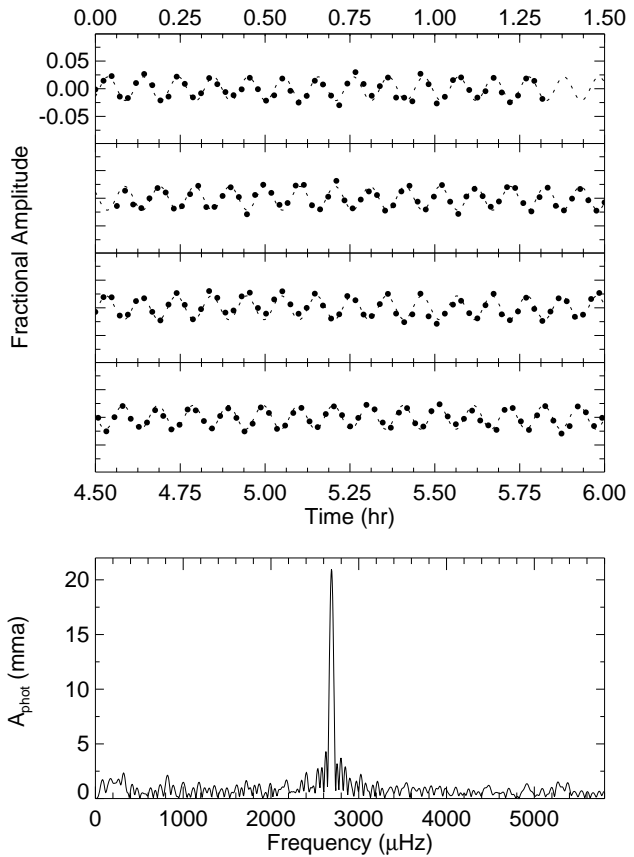}
  \caption{Light curve (top) produced from the spectrophotometry in the bandpass 3500-6200 \AA.  Although a parabolic fit has been removed to take out extinction variations, the curve has not been divided by that of a constant comparison star, attesting to the superb photometric quality of the night.  The amplitude spectrum (bottom) of the curve has a mean noise level of 0.65 mma.} 
  \label{fig:spec:lc}
\end{figure}

We created a light curve from our spectroscopic data by summing together the flux from all wavelength bins in each of the 248 spectra.  To correct for transparency variations, we attempted to divide the resulting light curve by that of the comparison star, but its low S/N ratio added noise to the light curve, and we abandoned the idea.  We did, however, correct for atmospheric extinction effects by fitting and normalising the light curve with a parabola.  We note that this normalisation removes any intrinsic variations on the order of the run length.  In spite of the absence of a comparison star by which to divide, the high photometric quality of the night and good S/N ratio in the spectra led to the superb light curve shown in Figure \ref{fig:spec:lc} above its FT.  

A least-squares fit to the curve reveals a single frequency of 2689.9 $\pm$ 0.6 $\mu$Hz (371.76 $\pm$ 0.09 s) with an amplitude of 21.5 $\pm$ 0.6 mma.  As with the time-series photometry, we correct for phase smearing, which leads to a damping of 7.45\%, and derive an amplitude of 23.3 $\pm$ 0.7 mma.  We detect no additional periodicities with signficant probabilities from the light curve, which had a mean noise level of 0.65 mma.  To compare the spectrophotometry to the PROMPT results, we convolved the spectra with the g' filter transmission function and repeated the light curve analysis.  We find a smearing-corrected amplitude of 23.2 $\pm$ 0.7 mma in agreement with the g' filter result from \S \ref{phot:analysis} .

\subsection{Radial velocity curve}
\label{rv_curve}
To construct a radial velocity curve, we individually fit the H$\beta$-H9 profiles in each spectrum with "pseudo" Lorentzian functions as described in \S \ref{mean_spectrum}, fixing the $\alpha$ parameter to the value determined from the average spectrum.  The individual spectra did not have the S/N ratio needed to fit the other absorption line profiles accurately, and, consequently, these lines were ignored in our analysis.  The locations of the profile centres in the mean spectrum were subtracted from those in the individual spectra, and the shifts from the mean were converted to km s$^{-1}$.  

\begin{figure}
  \centering
  \includegraphics{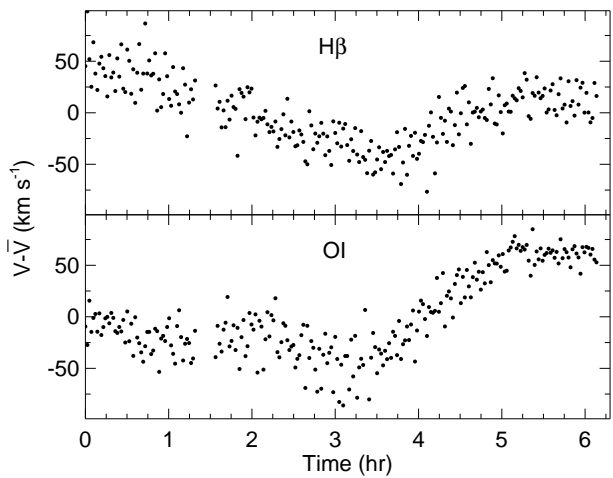}
   \caption{Radial velocities obtained from fitting pseudo Lorentzian profiles to the stellar H$\beta$ line (top) and the 5577 \AA\ O I sky emission line (bottom).  The large-scale variation is due to instrumental flexure.} 
  \label{fig:spec:rvc}
\end{figure}

The top panel in Figure \ref{fig:spec:rvc} shows the resulting radial velocity curve for H$\beta$, as an example.  The data exhibit a long-term variation spanning approximately 100 km s$^{-1}$ over the course of the run.  We attribute this structure to the combination of two factors: the global change in instrumental flexure and the movement of the star at the focal plane.  As the sky fills the entire slit, we may correct the former by examining the drift of atmospheric emission lines.  The latter problem is more difficult to correct, and, consequently, we do not attempt to do so.  Previous time-series photometry runs with the Goodman Spectrograph, however, show the drift due to guiding to be small, usually around 0.3\arcsec\ over a time span much longer than the period of CS 1246.

\begin{figure}
  \centering
  \includegraphics{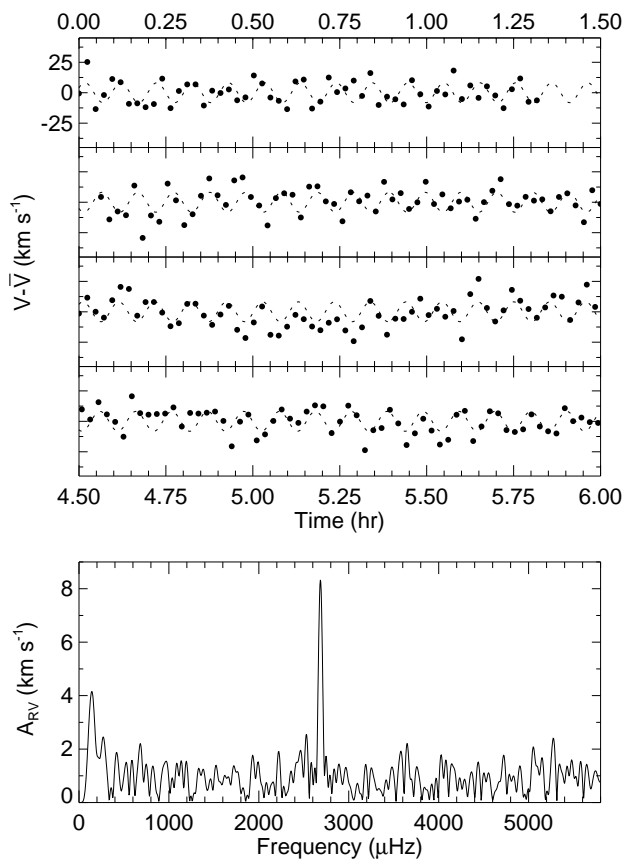}
   \caption{Flexure-corrected radial velocity curve (top) produced from the weighted average of the H$\beta$-H9 velocity curves.  The amplitude spectrum (bottom) has a mean noise level of 0.94 km s$^{-1}$.  Shortcomings in our flexure-correction are visible in the 3.25-4.0 hr section of the curve and give rise to the low-frequency peak in the FT near 143 $\mu$Hz.} 
  \label{fig:spec:rv}
\end{figure}

To compensate for flexure, we fit the O I sky line in each spectrum with a Gaussian using the GAUSSFIT routine in IDL and obtained the curve shown in the bottom panel of Figure \ref{fig:spec:rvc}.  A 5th-order polynomial matched the O I curve fairly well, and we removed this trend from the velocity curves of the stellar absorption lines.  After flexure correction, the velocity curves were linear with slopes dependent upon wavelength.  This result is expected from atmospheric refraction and allows us to approximate the guiding wavelength as 8400 \AA.  We computed the global radial velocity curve by removing the linear variations due to refraction and averaging the velocities from all of the H Balmer lines, weighting each curve by the inverse of its variance.  Figure \ref{fig:spec:rv} shows the resulting radial velocity curve along with its FT.

A strong radial velocity variation accompanies the photometric oscillation at the same frequency.  After barycentric correction of the velocities and the mid-exposure times, we find a semi-amplitude of 8.2 $\pm$ 1.0 km s$^{-1}$ by fitting a sine curve with frequency fixed to that derived from the PROMPT data.  Sine wave fits to the velocity curves of the individual H$\beta$-H9 lines give amplitudes of 8.7, 7.2, 7.5, 8.6, 8.9, and 9.8 km s$^{-1}$, respectively.  The uncertainty we report for the average velocity is the standard deviation of these values.  The increase in velocity with increasing Balmer order (with the exception of H$\beta$) implies the velocity is wavelength-dependent, as expected from weighting of the projected velocities over the observed hemisphere by limb-darkening.  Correcting for phase smearing, we derive a final semi-amplitude of 8.8 $\pm$ 1.1 km s$^{-1}$.  We claim no additional candidate frequencies after pre-whitening the data with the main pulsation period.  The large peak in the FT near 143 $\mu$Hz is a result of an imperfect flexure correction and should not be attributed to stellar oscillations.  The mean noise level in the amplitude spectrum is 0.94 km s$^{-1}$.

\subsection{Effective temperature and gravity variations}
\label{spec:temp_and_grav}

We determined T$_{eff}$ and log \textit{g} as a function of pulsation phase by phase-folding and averaging the spectra together in 10 evenly-spaced bins.  The resulting spectra had S/N ratios of approximately 240 pixel$^{-1}$ in the continuum near 5000 \AA.  We fit atmospheric models to the spectra in the manner described in \S \ref{mean_spectrum} to determine T$_{eff}$, log \textit{g}, and log \textit{N}(He)/\textit{N}(H) over a pulsation cycle.  As we observed no change in the He abundance over the 10 phase bins, we fixed the abundance to its average value and refit the other atmospheric parameters.

\begin{figure}
  \centering
  \includegraphics{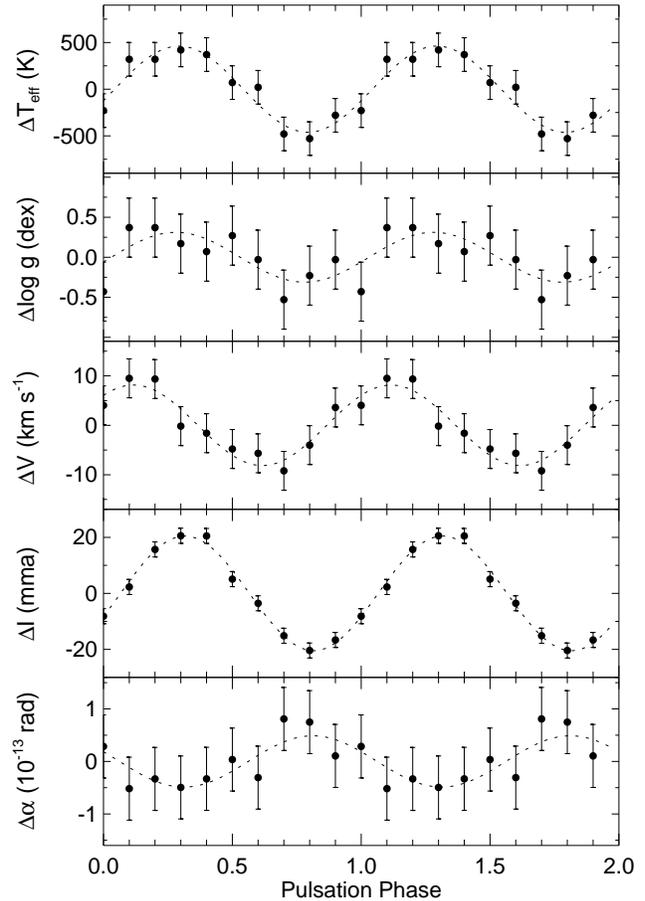}
   \caption{Variations in the effective temperature, surface gravity, radial velocity, spectrophotometry, and angular radius (from top to bottom) for the 371.7-s mode.  Non-linear, least-squares fits to the curves are shown as dotted lines, and the pulsation cycle is presented twice for visual clarity.  The curves shown are taken directly from the data and have \textit{not} been corrected for phase-smearing or projection effects.} 
  \label{fig:spec:variations}
\end{figure}

Figure \ref{fig:spec:variations} shows the T$_{eff}$ and log \textit{g} variations plotted above the radial velocity and light curves, which were phase-folded into the same 10 bins.  Since the temperature and gravity fits displayed sinusoidal variations, we fit sine waves to these curves using a $\chi^{2}$ fitting routine; the frequencies of the fits were fixed to the pulsational period.  We derive a temperature variation of 462 K about an equilibrium value of 28279 K and a gravity variation of 0.031 about a value of 5.44.  These amplitudes are subject to phase smearing from \textit{two} sources.  The finite integration time of the individual spectra in the series results in an underestimation of the amplitude by 7.45\%.  We must also consider the smearing that occurs when phase-folding and re-binning the spectra.  Each phased-binned spectrum has a pseudo-integration time of 37.17 s and leads to a further damping of 1.64\%.  Combining both smearing factors, we find that our fits to the T$_{eff}$ and log \textit{g} curves underestimate the true variations by 8.97\%.  Applying these corrections, we report semi-amplitudes of $\Delta$T = 507 $\pm$ 55 K and $\Delta$log \textit{g} = 0.034 $\pm$ 0.009.  The errors are derived using the deviations from the fits; they are more precise than those associated with the absolute temperature and gravity since systematic errors tend to cancel out in a differential analysis.

The phase relationships among the photometric, radial velocity, gravity, and temperature variations are apparent in Figure \ref{fig:spec:variations}.  If conditions are purely adiabatic, the temperature would be 90 $\deg$ out of phase with the radial velocity (180 $\deg$ out of phase with the radius).  We measure a difference of -72 $\pm$ 22 $\deg$.  Given the size of the error bars, we cannot claim deviations from adiabatic conditions using our current data set, although we do note this value is fairly deviant from 90 $\deg$.  The temperature, gravity, and flux variations are in phase to within the errors.

\subsection{Radial pulsation test and the stellar distance}
\label{spec:distance}
The simplicity of the light curve of CS 1246 combined with our high S/N spectra gives us the unique opportunity to check whether our observations are consistent with a radial oscillation.  The method we use is similar to that described by \citet{lyn84} and relies on the computation of the angular radius $\alpha$.  In \S \ref{spec:temp_and_grav} we found the best-fitting model spectrum to each of our observed phase-binned spectra, giving us the temperature $T$ at each phase $\phi$.  We take the monochromatic flux near 5500 \AA\ in the observed and model spectra for our calculations, as the continuum in this region is fairly smooth and free of absorption lines.  The model flux, $F_{\lambda}$, represents the flux at the photosphere and is a function of $T$ only, by definition.  The observed flux, $f_{\lambda}$, is a function of the temperature, the radius $R$, and the distance $d$.  Consequently, we can derive the angular radius $\alpha$ for each phase bin by computing the ratio of the model flux to the observed flux via the expression \begin{equation} \frac{F_{\lambda}(T)}{f_{\lambda}(T,R,d)} = \frac{d^{2}}{R^{2}}=\alpha^{-2}. \label{eqn:alpha}\end{equation}  Flux calibration errors, atmospheric extinction, and interstellar extinction will affect the observed flux; the latter two lead to an underestimation of $f_{\lambda}$.  We have already flux-calibrated the spectra and corrected them for atmospheric extinction in our reductions, but we have not yet considered the effects of interstellar extinction.  The temperature derived from our time-averaged spectrum implies a B-V colour of -0.29 $\pm$ 0.01.  \citet{rei88} reported a B-V colour of +0.28 $\pm$ 0.10, from which we compute the reddening as E(B-V) = 0.57 $\pm$ 0.11.  Assuming there is no companion star contributing to the reddening, we convert the E(B-V) estimate to an approximate extinction level of 1.8 mag for 5500 \AA\ using the results of \citet{sch98}.  Thus, our observed fluxes should be larger by a factor of 5.2 (=2.5$^{1.8}$).  Taking this factor into account, we computed the angular radius for each phase bin and derived the $\alpha$ variation shown in the bottom panel of Figure \ref{fig:spec:variations}.  A least-squares fit to the curve gives a semi-amplitude of $4.88*10^{-14}$ rad ($5.36*10^{-14}$ rad after smearing-correction) about a mean angular radius of $1.01*10^{-11}$ rad.  We stress again that errors in the flux calibration and extinction estimates lead to incorrect values for the angular radii and their relative differences.  

A plot of $\Delta\alpha$ vs $\Delta R$ will reveal a linear relationship for a radial pulsation mode, as shown by the relation \begin{equation} \Delta \alpha = \frac{\Delta R}{d}, \label{eqn:spec:dist}\end{equation} where the distance is the inverse of the slope.  Having already derived the $\alpha$ variations, we compute the variations in the physical radius by integrating the radial velocity curve.  We converted the observed expansion velocities into actual expansion velocities using a projection factor dependent upon relative limb darkening in the spectral lines.  \citet{mon01} computed such factors for early-type, radially pulsating stars, and their studies show a value of 1.4 to be appropriate for an sdB star like CS 1246.  By applying this factor, we assume from this point on that the 371.7-s mode is radial in nature and can only check whether our observations are consistent with this assumption.  The velocity amplitude becomes 12.3 $\pm$ 1.4 km s$^{-1}$ after application of the projection factor, and integration of this velocity gives a physical radius variation with semi-amplitude of 728 $\pm$ 83 km.  

\begin{figure}
  \centering
  \includegraphics{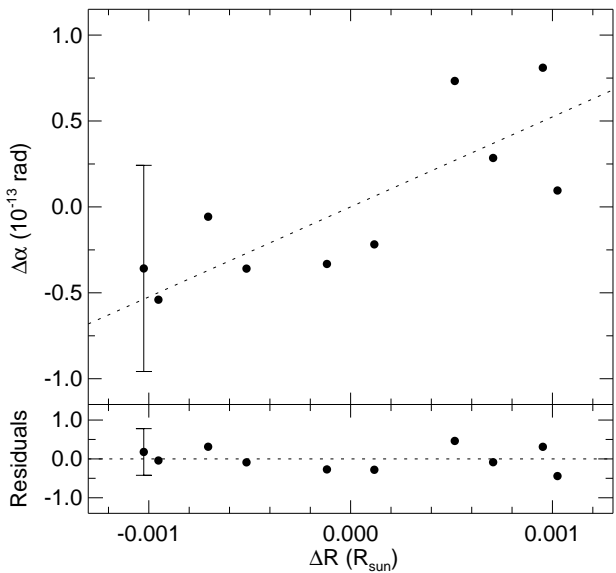}
   \caption{Plot of $\Delta \alpha$ vs $\Delta R$.  For a radial pulsator, these variables should exhibit a linear relationship.  The dashed line denotes the linear fit to these data.  The data exhibit a Pearson correlation coefficient of 0.78.  A representative error bar is shown for the leftmost data point and was computed by propagating errors through Eqn \ref{eqn:alpha}.}
  \label{fig:spec:distance}
\end{figure}

Figure \ref{fig:spec:distance} presents a plot of $\Delta \alpha$ against $\Delta$R.  To test for a linear relationship, we computed the Pearson correlation coefficient and find a value of 0.78.  The probability that 10 measurements of two uncorrelated variables can produce a correlation coefficient at or above this value is less than 2\% (see Appendix \ref{app:correlation}).  We can neither rule out the possibility that the 371.7-s mode in CS 1246 is a radial one, nor can we exclude the possibility of a mode with $\ell>0$.  

Still under the assumption of a radial mode, we fit a line to the data and determined the distance from the slope via Eqn. \ref{eqn:spec:dist}.  As $\Delta \alpha$ = 0 should correspond to the point where $\Delta$R = 0, we force the fit to pass through this point and derive a distance of d = 460 $\pm$ $^{190} _{100}$ pc.  An sdB star this close should be brighter than what \citet{rei88} measured.  In fact, with T$_{eff}$=28450$\pm$700 K and R=0.19$\pm$0.08 R$_{\sun}$ (see \S \ref{spec:radius_and_mass}), CS 1246 should have an apparent Visual magnitude of 12.7$\pm ^{0.9}_{0.5}$ when observed at the above distance.  This value implies a Visual extinction level of 1.9$\pm^{1.0}_{0.6}$ mags, in agreement with that found by \citet{rod60} and \citet{mat70} for the part of the Coalsack where CS 1246 resides.  

\subsection{The stellar radius and mass}
\label{spec:radius_and_mass}
We are now in the position to compute the radius and mass directly from our observations using two different methods.  The observed variations in log \textit{g} come about from the pulsational acceleration and changes in the radius, and after the former contribution is subtracted, the radius may be derived from the remaining difference.  Unfortunately, the uncertainty in $\Delta$log \textit{g} yields excessively large error bars on our determination of the radius and, consequently, the mass, and for this reason we deem this method unreliable with our current data set.  A second method proves to be more fruitful and is based upon the Baade-Wesselink method.  Any fractional change in the angular radius should yield the same fractional change in the physical radius.  Using this fact and linearizing Eqn. \ref{eqn:alpha}, we compute R from the following relation: \begin{equation} \frac{\Delta R}{R}=\frac{\Delta\alpha}{\alpha} = \frac{1}{2}\left(\frac{\Delta f_{\lambda}}{f_{\lambda}}-\frac{\Delta F_{\lambda}}{F_{\lambda}}\right). \label{spec:eqn:radius}\end{equation}  This expression yields a radius of R = 0.19 $\pm$ 0.08 R$_{\sun}$.  Similar results are obtained by calculating the radius directly from $\alpha$ and $d$, which gives R = 0.20 $\pm^{0.11}_{0.06}$ R$_{\sun}$.  We present the derivation of $R$ using Eqn \ref{spec:eqn:radius} to illustrate that errors in the observed flux cancel out and do \textit{not} affect the computation of the physical radius as they do the distance calculation.  

Combined with our measurement of the mean surface gravity, the radius derived from Eqn \ref{spec:eqn:radius} gives a mass of M = 0.39 $\pm^{0.30}_{0.13}$ M$_{\sun}$.  Again, the radius and mass we report are meaningless if the pulsation is non-radial.  Although our errors bars are large, the mass agrees with the commonly-accepted value of 0.5 M$_{\sun}$.  The radius is also consistent with those determined for other sdB stars from binary observations and from theoretical mass-radius relationships but has been computed using more direct observations.  Working backwards and assuming a canonical mass of 0.5 M$_{\sun}$, we would derive a radius of R = 0.22 R$_{\sun}$ from the mean log \textit{g} value.

\section{CONCLUSIONS}
We report the discovery of an exciting new member of the sdBV$_{r}$ class of pulsators, CS 1246.  Our simultaneous, multi-colour, time-series photometry reveals a single oscillation mode with period 371.7 s.  The amplitude of this mode depends strongly on wavelength and is largest in the u' filter.  We detect no additional periodicities in the frequency spectra, although we note our 0.41-m telescope detection limits are poor.  The light curve of CS 1246 appears to be the simplest of all sdBV stars.

Time-series spectroscopy proved to be a much more fruitful exercise than we originally anticipated.  We detected variations in temperature, gravity, and radial velocity at the same frequency as the photometric oscillations.  Interestingly, the ratios of these amplitudes (507 K, 0.034, 8.8 km s$^{-1}$) scale extremely well with those found by \citet{ost07} for the main mode of Balloon 090100001 (1186 K, 0.084, 18.9 km s$^{-1}$), which \citet{bar08} have confirmed as a radial oscillation.  By comparing changes in the angular and physical radii, we have shown the pulsation mode in CS 1246 is consistent with a radial one, but we cannot rule out non-radial modes with our current data set.  We were able, however, to calculate the radius and mass under the assumption of a radial pulsation, and find values consistent with those derived for other sdB stars.  Our study is the first to apply a method based on the Baade-Wesselink technique to an sdBV star.  There is room for improvement from long time-series spectroscopy runs, though, since the uncertainty in our mass does not allow one to distinguish between various sdB formation scenarios.  Better photometry and more precise radial velocities are required to achieve such a goal.  If the 371.7-s mode is confirmed to be radial, application of the Baade-Wesselink method to a more substantial data set could provide the most direct measurements of an sdB radius and mass ever made (except in cases where the sdBV is in an eclipsing binary system).

The atmospheric parameters we derived place CS 1246 near the boundary of the sdBV$_{r}$ and sdBV$_{s}$ instability strips, and, consequently, it could be a hybrid pulsator.  To date, three such sdBV$_{rs}$ stars have been noted in the literature:  Balloon 090100001 \citep{ore05,bar05,bar06}, HS 0702+6043 \citep{sch05,sch06,lut08}, and HS 2201+2610 \citep{lut09}.  Since studies of rapid and slow oscillations probe different depths in a star, the sdBV$_{rs}$ stars can provide much more information than pulsators exhibiting only one of these modes.  The amplitudes of slow oscillation modes tend to be small (a few mma), and we cannot rule out the possibility that CS 1246 is a hybrid pulsator since our detection limits in the low-frequency regime (2-3 mma) are comparable to this level.  A longer photometry run with a larger-aperture telescope is needed to search for slow pulsation modes.

The study of CS 1246 is complicated by the presence of the Coalsack Dark Nebula along our line of sight, which helps to explain two anomalous observations.  The first problem is the B-V colour reported by \citet{rei88}, which is too red for an sdB star.  Previous extinction studies of the Coalsack report reddening values similar to the one we find for CS 1246, so it is not impossible interstellar extinction is the only source of the observed reddening.  The Coalsack might also explain why our derived distance is less than what one would expect given the brightness of the star.  The effective temperature and radius we calculated imply an absolute Visual magnitude of 4.4 mag.  Combining this absolute magnitude with the apparent one, we derive a distance of nearly 1100 pc, much farther than the 460 pc calculated in \S\ref{spec:distance}.  Accounting for the approximately 1.8 mags of Visual extinction expected from the Coalsack, however, makes our computed distance consistent with the apparent magnitude.  We also note that at 460 pc, CS 1246 is located well behind the nebula, which has a distance of 174 pc \citep{rod60}.  If not for the Coalsack Dark Nebula, the star would be one of the brightest known sdBVs.

Finally, we draw attention to the potential CS 1246 has for the detection of its secular evolution and the discovery of orbiting low-mass companions.  Since the frequency spectrum is dominated by a single, large-amplitude oscillation, phase shifts should be relatively simple to detect through an O-C diagram.  The shifts may come about from secular evolution of the star as it evolves along or off of the extended horizontal branch or from wobbles due to orbital interactions with unseen, low-mass companions.  This method has already been used successfully to find a planet around the sdB star V391 Pegasi \citep{sil07}.  As PROMPT is an ideal tool for such a study, we are currently building up an ephemeris for CS 1246 and look forward to producing our first O-C diagram in the future.

\section*{Acknowledgments}

We acknowledge the support of the National Science Foundation, under award AST-0707381 and are grateful to the Abraham Goodman family for the financial support that made the spectrograph a reality.  We recognize the observational support of Patricio Ugarte and Alberto Pasten at the SOAR telescope.  BNB would personally like to thank John Tumbleston for help during the analysis and Chris Koen and Simon Jeffery for useful discussions during the fourth sdOB meeting in Shanghai, China.  Finally, we thank an anonymous referee for helping us clarify our presentation in various parts of this manuscript.

\appendix

\section{Correlation Coefficient Probabilities}
\label{app:correlation}

We compute the correlation coefficient and corresponding probabilities using the methods described by \citet{tay97}.  The linear correlation coefficient $r$ between $N$ points (x$_{i}$,y$_{i}$) can be computed using the expression \begin{equation} r=\frac{\Sigma(x_{i} - \bar{x})(y_{i} - \bar{y})}{\sqrt{\Sigma(x_{i} - \bar{x})^{2}(y_{i} - \bar{y})^{2}}}.\end{equation}  The coefficient $r$ will be close to $\pm$ 1 if the linear correlation is strong and near 0 if the points are uncorrelated.  As even uncorrelated points can produce a non-zero coefficient, one must further quantify the significance of $r$ by considering the number of measurements.  The probability that $N$ measurements of two uncorrelated variables result in a coefficient greater than or equal to $r_{o}$ is given by \begin{equation} P_{N}(\left |r\right | \geq \left |r_{o}\right |) = \frac{2\Gamma[(N-1)/2]}{\sqrt{\pi}\Gamma[(N-2)/2]}\int^{1}_{\left |r_{o}\right |} (1-r^{2})^{(N-4)/2}dr.\end{equation}  This probability strongly depends upon the number of measurements $N$; the probability that two uncorrelated variables give a set correlation coefficient decreases with increasing $N$.  

In our case, the $\Delta\alpha$ and $\Delta R$ points measured give a coefficient of 0.78.  The above expression shows there is less than a 2\% chance that two uncorrelated variables give a coefficient at or above 0.78.  Thus, we have significant but not definitive evidence of a linear correlation.

\bsp


\begin{thebibliography}{}

\bibitem[\protect\citeauthoryear{{Baldry}}{{Baldry}}{1999}]{bal99}
{Baldry} I.,  1999, PhD thesis, University of Sydney

\bibitem[\protect\citeauthoryear{{Baran}, {Oreiro}, {Pigulski}, {P{\'e}rez
  Hern{\'a}ndez} \& {Ulla}}{{Baran} et~al.}{2006}]{bar06}
{Baran} A. et al.,  2006, Baltic Astronomy, 15, 227

\bibitem[\protect\citeauthoryear{{Baran}, {Pigulski}, {Kozie{\l}}, {Og{\l}oza},
  {Silvotti} \& {Zo{\l}a}}{{Baran} et~al.}{2005}]{bar05}
{Baran} A. et al.,  2005, MNRAS, 360, 737

\bibitem[\protect\citeauthoryear{{Baran}, {Pigulski} \& {O'Toole}}{{Baran}
  et~al.}{2008}]{bar08}
{Baran} A., {Pigulski} A., {O'Toole} S.~J.,  2008, MNRAS, 385, 255

\bibitem[\protect\citeauthoryear{{Bergeron}, {Wesemael}, {Lamontagne},
  {Fontaine}, {Saffer} \& {Allard}}{{Bergeron} et~al.}{1995}]{ber95}
{Bergeron} P. et al.,  1995, ApJ, 449, 258

\bibitem[\protect\citeauthoryear{{Brassard}, {Fontaine}, {Bill{\`e}res},
  {Charpinet}, {Liebert} \& {Saffer}}{{Brassard} et~al.}{2001}]{bra01}
{Brassard} P. et al.,  2001, ApJ, 563, 1013

\bibitem[\protect\citeauthoryear{{Charpinet}, {Fontaine}, {Brassard},
  {Bill{\`e}res}, {Green} \& {Chayer}}{{Charpinet} et~al.}{2005}]{cha05}
{Charpinet} S. et al.,  2005, A\&A, 443, 251

\bibitem[\protect\citeauthoryear{{Charpinet}, {Fontaine}, {Brassard}, {Chayer},
  {Rogers}, {Iglesias} \& {Dorman}}{{Charpinet} et~al.}{1997}]{cha97}
{Charpinet} S. et al.,  1997, ApJL, 483, L123

\bibitem[\protect\citeauthoryear{{Charpinet}, {Fontaine}, {Brassard} \&
  {Dorman}}{{Charpinet} et~al.}{1996}]{cha96}
{Charpinet} S., {Fontaine} G., {Brassard} P., {Dorman} B.,  1996, ApJL,
  471, L103

\bibitem[\protect\citeauthoryear{{Clemens}, {Crain} \& {Anderson}}{{Clemens}
  et~al.}{2004}]{cle04}
{Clemens} J.~C.,  {Crain} J.~A.,    {Anderson} R.,  2004, in {Moorwood}
  A.~F.~M.,  {Iye} M.,  eds, SPIE Conference Series Vol. 5492 of SPIE Conference Series, {The Goodman
  spectrograph}.
pp 331--340

\bibitem[\protect\citeauthoryear{{Crawford}}{{Crawford}}{1991}]{cra91}
{Crawford} I.~A.,  1991, AAP, 246, 210

\bibitem[\protect\citeauthoryear{{D'Cruz}, {Dorman}, {Rood} \&
  {O'Connell}}{{D'Cruz} et~al.}{1996}]{dcr96}
{D'Cruz} N.~L.,  {Dorman} B.,  {Rood} R.~T.,    {O'Connell} R.~W.,  1996, ApJ,
  466, 359

\bibitem[\protect\citeauthoryear{{Dorman}, {Rood} \& {O'Connell}}{{Dorman}
  et~al.}{1993}]{dor93}
{Dorman} B.,  {Rood} R.~T.,    {O'Connell} R.~W.,  1993, ApJ, 419, 596

\bibitem[\protect\citeauthoryear{{Drechsel}, {Heber}, {Napiwotzki},
  {{\O}stensen}, {Solheim}, {Johannessen}, {Schuh}, {Deetjen} \&
  {Zola}}{{Drechsel} et~al.}{2001}]{dre01}
{Drechsel} H. et al.,  2001,  AAP, 379, 893

\bibitem[\protect\citeauthoryear{{Franco}}{{Franco}}{1989}]{fra89a}
{Franco} G.~A.~P.,  1989, AAP, 215, 119

\bibitem[\protect\citeauthoryear{{Franco}}{{Franco}}{2000}]{fra00}
{Franco} G.~A.~P.,  2000, MNRAS, 315, 611

\bibitem[\protect\citeauthoryear{{Han}, {Podsiadlowski} \& {Lynas-Gray}}{{Han}
  et~al.}{2007}]{han07}
{Han} Z.,  {Podsiadlowski} P.,    {Lynas-Gray} A.~E.,  2007, MNRAS, 380, 1098

\bibitem[\protect\citeauthoryear{{Han}, {Podsiadlowski}, {Maxted} \&
  {Marsh}}{{Han} et~al.}{2003}]{han03}
{Han} Z.,  {Podsiadlowski} P.,  {Maxted} P.~F.~L.,    {Marsh} T.~R.,  2003,
  MNRAS, 341, 669

\bibitem[\protect\citeauthoryear{{Han}, {Podsiadlowski}, {Maxted}, {Marsh} \&
  {Ivanova}}{{Han} et~al.}{2002}]{han02}
{Han} Z. et al.,  2002, MNRAS, 336, 449

\bibitem[\protect\citeauthoryear{{Heber}}{{Heber}}{1986}]{heb86}
{Heber} U.,  1986, AAP, 155, 33

\bibitem[\protect\citeauthoryear{{Heber}}{{Heber}}{2009}]{heb09}
{Heber} U.,  2009, ARAA, 47, 211

\bibitem[\protect\citeauthoryear{{Horne} \& {Baliunas}}{{Horne} \&
  {Baliunas}}{1986}]{hor86}
{Horne} J.~H.,  {Baliunas} S.~L.,  1986, ApJ, 302, 757

\bibitem[\protect\citeauthoryear{{Kilkenny}, {Koen}, {O'Donoghue} \&
  {Stobie}}{{Kilkenny} et~al.}{1997}]{kil97}
{Kilkenny} D.,  {Koen} C.,  {O'Donoghue} D.,    {Stobie} R.~S.,  1997, MNRAS,
  285, 640

\bibitem[\protect\citeauthoryear{{Lutz}, {Schuh}, {Silvotti}, {Bernabei},
  {Dreizler}, {Stahn} \& {Huegelmeyer}}{{Lutz} et~al.}{2009}]{lut09}
{Lutz} R. et al.,  2009, ArXiv e-prints

\bibitem[\protect\citeauthoryear{{Lutz}, {Schuh}, {Silvotti}, {Dreizler},
  {Green}, {Fontaine}, {Stahn}, {H{\"u}gelmeyer} \& {Husser}}{{Lutz}
  et~al.}{2008}]{lut08}
{Lutz} R. et al.,  eds, Hot Subdwarf
  Stars and Related Objects Vol.~392 of Astronomical Society of the Pacific
  Conference Series, {Light Curve Analysis of the Hybrid SdB PulsatorsHS
  0702+6043 and HS 2201+2610}.
pp 339-342

\bibitem[\protect\citeauthoryear{{Lynas-Gray}, {Schoenberner}, {Hill} \&
  {Heber}}{{Lynas-Gray} et~al.}{1984}]{lyn84}
{Lynas-Gray} A.~E.,  {Sch{\"o}nberner} D.,  {Hill} P.~W.,    {Heber} U.,  1984,
  MNRAS, 209, 387

\bibitem[\protect\citeauthoryear{{Markwardt}}{{Markwardt}}{2009}]{mar09}
{Markwardt} C.~B., 2009, ASP Conf. Ser. 411, Astronomical Data Analysis Software and Systems XVIII, ed. D. Bohlender, D. Durand, \& P. Dowler (San Francisco, CA: ASP), 251 

\bibitem[\protect\citeauthoryear{{Mattila}}{{Mattila}}{1970}]{mat70}
{Mattila} K.,  1970, AAP, 8, 273

\bibitem[\protect\citeauthoryear{{Maxted}, {Heber}, {Marsh} \&
  {North}}{{Maxted} et~al.}{2001}]{max01}
{Maxted} P.~f.~L.,  {Heber} U.,  {Marsh} T.~R.,    {North} R.~C.,  2001,
  MNRAS, 326, 1391

\bibitem[\protect\citeauthoryear{{Monta{\~n}{\'e}s Rodriguez} \&
  {Jeffery}}{{Monta{\~n}{\'e}s Rodriguez} \& {Jeffery}}{2001}]{mon01}
{Monta{\~n}{\'e}s Rodriguez} P.,  {Jeffery} C.~S.,  2001, AAP, 375, 411

\bibitem[\protect\citeauthoryear{{Montgomery} \& {O'Donoghue}}{{Montgomery} \&
  {O'Donoghue}}{1999}]{mon99}
{Montgomery} M.~H.,  {O'Donoghue} D.,  1999, Delta Scuti Star Newsletter, 13,
  28

\bibitem[\protect\citeauthoryear{{Oreiro}, {P{\'e}rez Hern{\'a}ndez}, {Ulla},
  {Garrido}, {{\O}stensen} \& {MacDonald}}{{Oreiro} et~al.}{2005}]{ore05}
{Oreiro} R. et al.,  2005, AAP, 438, 257

\bibitem[\protect\citeauthoryear{{{\O}stensen}, {Telting} \&
  {Heber}}{{{\O}stensen} et~al.}{2007}]{ost07}
{{\O}stensen} R.,  {Telting} J.,    {Heber} U.,  2007, Communications in
  Asteroseismology, 150, 265

\bibitem[\protect\citeauthoryear{{{\O}stensen}}{{{\O}stensen}}{2006}]{ost06}
{{\O}stensen} R.~H.,  2006, Baltic Astronomy, 15, 85

\bibitem[\protect\citeauthoryear{{{\O}stensen}}{{{\O}stensen}}{2009}]{ost09}
{{\O}stensen} R.~H.,  2009, Communications in Asteroseismology, 159, 75

\bibitem[\protect\citeauthoryear{{O'Toole}, {Bedding}, {Kjeldsen}, {Teixeira},
  {Roberts}, {van Wyk}, {Kilkenny}, {D'Cruz} \& {Baldry}}{{O'Toole}
  et~al.}{2000}]{oto00}
{O'Toole} S.~J.,  {Bedding} T.~R.,  {Kjeldsen} H.,  {Teixeira} T.~C.,
  {Roberts} G.,  {van Wyk} F.,  {Kilkenny} D.,  {D'Cruz} N.,    {Baldry} I.~K.,
   2000, ApJl, 537, L53

\bibitem[\protect\citeauthoryear{{Press}, {Teukolsky}, {Vetterling} \&
  {Flannery}}{{Press} et~al.}{2007}]{pre07}
{Press} W.~H.,  {Teukolsky} S.~A.,  {Vetterling} W.~T.,    {Flannery} B.~P.,
  2007, {Numerical Recipes 3rd Edition: The Art of Scientific Computing}.
Cambridge: University Press, |c2007, 3rd ed.

\bibitem[\protect\citeauthoryear{{Randall}, {Green}, {van Grootel}, {Fontaine},
  {Charpinet}, {Lesser}, {Brassard}, {Sugimoto}, {Chayer}, {Fay}, {Wroblewski},
  {Daniel}, {Story} \& {Fitzgerald}}{{Randall} et~al.}{2007}]{ran07}
{Randall} S.~K.,  {Green} E.~M.,  {van Grootel} V.,  {Fontaine} G.,
  {Charpinet} S.,  {Lesser} M.,  {Brassard} P.,  {Sugimoto} T.,  {Chayer} P.,
  {Fay} A.,  {Wroblewski} P.,  {Daniel} M.,  {Story} S.,    {Fitzgerald} T.,
  2007, AAP, 476, 1317

\bibitem[\protect\citeauthoryear{{Reichart}}{{Reichart} et~al.}{2009}]{rei09}
{Reichart} D.E. et al., 2009, Nuovo Cimento C Geophysics Space Physics C, 28, 767.

\bibitem[\protect\citeauthoryear{{Reid}, {Wegner}, {Wickramasinghe} \&
  {Bessell}}{{Reid} et~al.}{1988}]{rei88}
{Reid} N.,  {Wegner} G.,  {Wickramasinghe} D.~T.,    {Bessell} M.~S.,  1988,
  AJ, 96, 275

\bibitem[\protect\citeauthoryear{{Rodgers}}{{Rodgers}}{1960}]{rod60}
{Rodgers} A.~W.,  1960, MNRAS, 120, 163

\bibitem[\protect\citeauthoryear{{Saffer}, {Green} \& {Bowers}}{{Saffer}
  et~al.}{2001}]{saf01}
{Saffer} R.~A.,  {Green} E.~M.,    {Bowers} T.,  2001, in {J.~L.~Provencal,
  H.~L.~Shipman, J.~MacDonald, \& S.~Goodchild } ed., 12th European Workshop on
  White Dwarfs Vol.~226 of Astronomical Society of the Pacific Conference
  Series, {The Binary Origins Of Hot Subdwarfs: New Radial Velocities}.
pp 408--+

\bibitem[\protect\citeauthoryear{{Saffer}, {Liebert} \& {Olszewski}}{{Saffer}
  et~al.}{1988}]{saf88}
{Saffer} R.~A.,  {Liebert} J.,    {Olszewski} E.~W.,  1988, ApJ, 334, 947

\bibitem[\protect\citeauthoryear{{Schlegel}, {Finkbeiner} \&
  {Davis}}{{Schlegel} et~al.}{1998}]{sch98}
{Schlegel} D.~J.,  {Finkbeiner} D.~P.,    {Davis} M.,  1998, ApJ, 500, 525

\bibitem[\protect\citeauthoryear{{Schuh}, {Huber}, {Dreizler}, {Heber},
  {O'Toole}, {Green} \& {Fontaine}}{{Schuh} et~al.}{2006}]{sch06}
{Schuh} S. et al., 2006, AAP, 445, L31

\bibitem[\protect\citeauthoryear{{Schuh}, {Huber}, {Green}, {O'Toole},
  {Dreizler}, {Heber} \& {Fontaine}}{{Schuh} et~al.}{2005}]{sch05}
{Schuh} S. et al., 2005, in Koester D., Moehler S., eds, ASP Conf.Ser. Vol. 334,14th European Workshop on White Dwarfs, Discovery of a Long-Period Photometric   Variation in the V361 Hya Star HS 0702+6043. Astron. Soc. Pac., San Francisco, P. 530.

\bibitem[\protect\citeauthoryear{{Silvotti}, {Schuh}, {Janulis}, {Solheim},
  {Bernabei} \& {{\O}stensen}}{{Silvotti} et~al.}{2007}]{sil07}
{Silvotti} R. et al., 2007, Nature, 449, 189

\bibitem[\protect\citeauthoryear{{Taylor}}{{Taylor}}{1997}]{tay97}
{Taylor} J.,  1997, {Introduction to Error Analysis, the Study of Uncertainties
  in Physical Measurements, 2nd Edition}.
University Science Books

\bibitem[\protect\citeauthoryear{{Telting} \& {{\O}stensen}}{{Telting} \&
  {{\O}stensen}}{2004}]{tel04}
{Telting} J.~H.,  {{\O}stensen} R.~H.,  2004, AAP, 419, 685

\bibitem[\protect\citeauthoryear{{Tremblay}, {Fontaine}, {Brassard}, {Bergeron}
  \& {Randall}}{{Tremblay} et~al.}{2006}]{tre06}
{Tremblay} P.-E.,  {Fontaine} G.,  {Brassard} P.,  {Bergeron} P.,    {Randall}
  S.~K.,  2006, ApJS, 165, 551

\bibitem[\protect\citeauthoryear{{Thompson} \& {Mullally}}{{Thompson} \&
  {Mullally}}{2009}]{tho09}
{Thompson} S.~E.,  {Mullally} F.,  2009, Journal of Physics Conference Series,
  172, 012081

\bibitem[\protect\citeauthoryear{{Tillich}, {Heber}, {O'Toole}, {{\O}stensen}
  \& {Schuh}}{{Tillich} et~al.}{2007}]{til07}
{Tillich} A.,  {Heber} U.,  {O'Toole} S.~J.,  {{\O}stensen} R.,    {Schuh} S.,
  2007, AAP, 473, 219

\bibitem[\protect\citeauthoryear{{Wood}, {Zhang} \& {Robinson}}{{Wood}
  et~al.}{1993}]{woo93}
{Wood} J.~H.,  {Zhang} E.-H.,    {Robinson} E.~L.,  1993, MNRAS, 261, 103

\bibitem[\protect\citeauthoryear{{Zhang}, {Chen} \& {Han}}{{Zhang}
  et~al.}{2009}]{zha09}
{Zhang} X.,  {Chen} X.,    {Han} Z.,  2009, AAP, 504, L13

\end{thebibliography}
\end{document}